\documentclass{article}
\def\giorno{12 August  2004}

\def\.#1{\dot #1}
\def\D{{\cal D}}
\def\G{{\cal G}}
\def\H{{\cal H}}
\def\I{{\cal I}}

\def\L{{\cal L}}
\def\M{{\cal M}}
\def\R{{\bf R}}  %% reals
\def\S{{\cal S}}
\def\T{{\rm T}}
\def\V{{\cal V}}
\def\PT{\~\Phi}
\def\XT{\~X}

\def\sse{\subseteq}
\def\ss{\subset}
\def\pa{\partial}

\def\=#1{{\widetilde #1}}
\def\~#1{\widetilde #1}
\def\.#1{\dot #1}
\def\^#1{\widehat #1}

\def\sse{\subseteq}

\def\d{{\rm d}}       %% derivative

\def\({\left(}
\def\){\right)}
\def\[{\left[}
\def\]{\right]}

\def\a{\alpha}

\def\b{\beta}
\def\be{\beta}

\def\ga{\gamma}
\def\de{\delta}   %% NON ridefinire come \d !!!!

\def\phi{\varphi}
\def\la{\lambda}
\def\La{\Lambda}
\def\s{\sigma}
\def\om{\omega}
\def\vth{\vartheta}

\def\phi{\varphi}
\def\Ga{\Gamma}
\def\De{\Delta}

\def\w{\wedge}

\def\interno{\hskip 2pt \vbox{\hbox{\vbox to .18 truecm{\vfill\hbox to .25 truecm {\hfill\hfill}\vfill}\vrule}\hrule}\hskip 2 pt}

\def\EOP{~\hfill $\diamondsuit$} % end of proof
\def\EOD{~\hfill $\clubsuit$}    % end of definition
\def\EOR{~\hfill $\odot$}        % end of remark

\begin{document}

\title{\bf On the relation between standard and $\mu$-symmetries for PDEs}

\author{Giampaolo Cicogna\footnote{Dipartimento  di Fisica ``E. Fermi'', Universit\`a di Pisa, and INFN Sez. di Pisa, Largo B. Pontecorvo 2, I--56127 Pisa (Italy); e-mail: cicogna@df.unipi.it}, Giuseppe Gaeta\footnote{Dipartimento di Matematica, Universit\`a di Milano, via Saldini 50, I--20133 Milano (Italy); e-mail: gaeta@mat.unimi.it}, and Paola Morando\footnote{Dipartimento di Matematica, Politecnico di Torino, Corso Duca degli Abruzzi 24, I--10129 Torino (Italy); e-mail: paola.morando@polito.it} }

\date{ \giorno  }

\maketitle

\noindent {\bf Summary.} 
We give a geometrical interpretation of the notion of $\mu$-prolongations of vector fields and of the related concept of $\mu$-symmetry for partial differential equations (extending to PDEs the notion of $\lambda$-symmetry for ODEs). We give in particular a result concerning the relationship between $\mu$-symmetries and standard exact symmetries. The notion is also extended to the case of conditional and partial symmetries, and we analyze the relation between local $\mu$-symmetries and nonlocal standard symmetries.

\par\noindent

{\tt PACS: 02.20.-a ; 02.30.Jr \ . \ MSC: 58J70 ; 35A30 \ . }

\section*{Introduction}

Symmetry analysis and reduction methods are well established tools to study nonlinear differential equations, see e.g. \cite{Gae,Kra,Olv1,Olv2,Ste,Win}. It was recently pointed out that the class of transformations which are useful for the symmetry reduction -- and for finding solutions -- of ODEs is not limited to (Lie-point, generalized, non-local...) proper symmetries, but extends to a wider class of transformations, which were christened ${\cal C}^\infty$-symmetries or $\la$-symmetries, as they depend on a ${\cal C}^\infty$ function $\la$ \cite{MuR1}. (See \cite{GMM,MuR2} for some applications).

It is remarkable that for such transformations one can perform a ``symmetry reduction'' in exactly the same way as for standard symmetries; transformations enjoying this property were then studied in full generality -- for scalar ODEs~-- and it was shown that standard and $\la$-symmetries are essentially the only possibility, up to considering contact version of both \cite{Sac}.

The concept of $\lambda$-symmetries was extended to the PDE frame (both scalar equations and systems) with $p$ independent variables $x^i$ and $q$ dependent ones $u^a$ in \cite{GaMo}; in order to do this, the central object is a horizontal one-form $\mu = \La_i \d x^i$, where $\La_i (x,u^{(n)}) \in g \ell (q)$, and thus one speaks of $\mu$-symmetries. We stress that $\mu$ is not entirely arbitrary: indeed its coefficients must satisfy a compatibility condition, which reads $ D_i \La_j - D_j \La_i + [\La_i , \La_j] = 0 $ for all $i,j = 1,...,p$ (note this is automatically satisfied for $p=1$, and takes a simpler form for $q=1$).

This means that $\mu$ satisfies the horizontal Maurer-Cartan equation $D \mu + (1/2) [\mu , \mu] = 0$, with $D$ the total horizontal external derivative operator (defined on functions as $D f = (D_i f) \d x^i$, with $D_i$ the total derivative with respect to $x^i$, and accordingly on forms; see e.g. \cite{Olv1,Olv2} for details). Equivalently, the (matrix) operators $\nabla_i := D_i + \La_i$ commute, $[\nabla_i , \nabla_j ] = 0 $.

It should be stressed that $\mu$-symmetries are {\it not} symmetries in proper sense \cite{Gae,Olv1,Olv2,Ste}, i.e. they do not, in general, map solutions into solutions; nevertheless, it was shown in \cite{GaMo} that they can be used to perform ``symmetry reduction'' of PDEs and systems of PDEs -- that is, to obtain invariant solutions~-- via exactly the same method used for standard symmetries.

%\medskip

The main goal of this note is to give a geometrical interpretation of the concept of $\mu$-prolongations, and hence of $\mu$-symmetries.

These concepts will be recalled, together with some related results which are relevant in the  following, in section 1. The geometrical interpretation mentioned above will be provided in section 2 in the form of a result describing the relationship between $\mu$-prolongations and standard prolongations, and hence of $\mu$-symmetries and standard symmetries. With this interpretation we will also be able to extend to the framework of $\mu$-symmetries the concepts of conditional \cite{CGspri,Gae,LeWi,Win} and partial \cite{CGpar} symmetries, in section  3. Starting again from the interpretation obtained in section 2 we also study, in section 4, the interrelation between local $\mu$-symmetries and nonlocal standard symmetries (of exponential type). After a short discussion on our approach and results in section 5, several examples are considered in section 6.

Finally, we would like to remark that the geometrical characterization of $\mu$-prolongations given below only applies in full when there are at least two independent variables, i.e. the ODE case is in this respect a strongly degenerate one (see remark 7). This also explains why such an interpretation -- and generalization to conditional and partial symmetries -- could not be obtained in the seemingly simpler ODE case.

\bigskip

\noindent {\bf Acknowledgements.} This work was partially supported by GNFM-INdAM through the project {\it ``Simmetrie e tecniche di riduzione per equazioni differenziali di interesse fisico-matematico''}.

\section{Prolongations, $\mu$-prolongations, and \\ $\mu$-symmetries}

In this section we fix notation and collect some background material, either standard \cite{Kra,Olv1,Ste} or specific to $\mu$-symmetries \cite{GaMo}, of use in the following.

We will denote by $M$ the space of independent and dependent variables; these will be denoted respectively as $x \in B$ and $u \in F$, with  $x = (x^1 , ... , x^p )$ and $u = (u^1,...,u^q)$ in local coordinates.

Thus $M$ is the total space of a fiber bundle $(M,\pi,B)$ over $B$, with fiber $\pi^{-1} (x) = F$. In the standard case, $B \sse \R^p $, $F = \R^q$, and $M = B \times \R^q$; in this note we will mainly stick to this setting, for ease of discussion.

\subsection{Prolongations and $\mu$-prolongations}

The bundle $M$ can be prolonged to the $n$-th jet bundle $(J^{(n)} M , \pi_n , B)$; the local coordinates $(x,u)$ in $M$ naturally induce local coordinates $(x,u^{(n)} )$ in $J^{(n)} M$, given by $u^a_J$, where $J = (j_1 , ... , j_p)$ is a multiindex of order $|J| := j_1 + ... + j_p$ and $u^a_J = \pa^{|J|} u / [(\pa x^1)^{j_1} ... (\pa x^p)^{j_p} ]$. We also use the notation $u_{J,i} := \pa u^a_J / \pa x^i$.

The jet space $J^{(n)} M$ is naturally endowed with a contact structure, spanned by the set of $F$-valued contact one-forms $\vth_J$ with $|J| < n$; the reader is referred to \cite{Mal,Str} for vector-valued forms. In the local coordinates introduced above the contact forms are given by $\vth_J = (\vth^1_J , ... , \vth^q_J )$, with 
$$ \vth^a_J \ := \ \d u^a_J \ - \ u^a_{J,m} \, \d x^m \ . \eqno(1.1) $$ 
The contact structure $\Theta$ is the $C^\infty [J^{(n)} M]$ module spanned by the $\vth_J$, $|J| < n$.

A Lie-point vector field $X$ in $M$ is naturally prolonged to a vector field $X^{(n)}$ in $J^{(n)} M$. This is the unique vector field which projects to $X$ when restricted to $M$, and which preserves the contact structure. The latter condition means that for any $\vth \in \Theta$, there is a (possibly zero) $\^\vth \in \Theta $ such that $ \L_X (\vth ) = \^\vth$.

If $X$ and its prolongation $X^{(n)}$ are written in local coordinates as
$$ \begin{array}{ll}
X    &=  \ \ \xi^i \, (\pa / \pa x^i) \ + \ \phi^a \, (\pa / \pa u^a) \ , \\
X^{(n)}   &=  \ \ X \ + \ \Phi^a_J \, (\pa / \pa u^a_J) \ , \end{array} \eqno(1.2) $$ 
then the coefficients $\Phi^a_J$ satisfy the {\it prolongation formula} (with $\Phi^a_0 \equiv \phi^a$) 
$$ \Phi^a_{J,i} \ = \ D_i \Phi^a_J \ - \ u^a_{J,m} D_i \xi^m \ . \eqno(1.3) $$
In the case of $\mu$-prolongations, we equip $M$ with a horizontal one-form $\mu = \La_i \d x^i$ with values in the algebra $\G := g \ell (q)$; this is the Lie algebra of the group $G := GL (q)$ of $q$-dimensional nonsingular matrices (we refer again to \cite{Mal,Str} for matrix-valued forms). That is, the $\La_i$ are matrix functions defined on a suitable jet space $J^{(k)} M$ (if $k \le 1$ we will have proper vector fields in each jet space, otherwise we deal with generalized vector fields).

Given the (possibly, generalized) vector field $X$, its $\mu$-prolongation is the unique vector field which projects to $X$ when restricted to $M$, and which ``$\mu$-preserves'' (see \cite{GaMo}) the contact structure $\Theta$. By this we mean that for any $\vth \in \Theta$ there is a (possibly zero) $\^\vth \in \Theta$ such that $ \L_Y (\vth^a ) + ( Y \interno [ (\La_i)^a_b \vth^b ] ) \d x^i = \^\vth^a$.

If $X$ is written in local coordinates as in (1.2), and its $\mu$-prolongation as
$$ Y \ := \ X \ + \ \Psi^a_J \, (\pa / \pa u^a_J) \ , \eqno(1.4) $$
then the coefficients $\Psi^a_J$ satisfy the {\it $\mu$-prolongation formula}
$$ \Psi^a_{J,i} \ = \ (\nabla_i)^a_b \, \Psi^b_J \ - \ u^b_{J,m} \, (\nabla_i )^a_b \, \xi^m \ . \eqno(1.5) $$ 
Here $\nabla_i$ is the matrix operator defined by $\nabla_i = D_i + \La_i$, i.e.
$$ (\nabla_i)^a_b \ := \ \de^a_b \, D_i \ + \ (\La_i )^a_b \ . \eqno(1.6) $$

Note that, as discussed in \cite{GaMo}, $Y$ is well defined by (1.5) if and only if the $\La_i$ satisfy the compatibility condition 
$$ D_i \La_j \ - \ D_j \La_i \ + \ [ \La_i , \La_j ] \ = \ 0 \ , \eqno(1.7') $$ 
which is equivalently written as 
$$ \[ \, \nabla_i \, , \, \nabla_j \, \] \ = \ 0 \ . \eqno(1.7'') $$
If (1.7) is satisfied only on a submanifold $S \ss J^{(n)} M$, then $Y$ is properly defined by (1.5) only on $S$; in case $Y : S \to \T S$ (i.e. $S$ is $Y$-invariant), it still makes sense to study the $Y$-action on $S$, see remark 2 below.

\medskip\noindent 
{\bf Remark 1.} The coefficients of the $\mu$-prolongation $Y$ can also be described in terms of the coefficients of the ordinary prolongation $X^{(n)}$ of the same vector field $X$. If we write $\Psi^a_J = \Phi^a_J + F^a_J$, the difference terms $F^a_J$ satisfy (with of course $F^a_0 = 0$) the recursion relations $F^a_{J,i} = [ \de^a_b D_i + (\La_i)^a_b ] F^b_J  + (\La_i)^a_b D_J Q^b$, where $Q$ is the characteristic vector for $X$, i.e. $Q^a := \phi^a - u^a_i \xi^i$.

It follows from this that $\Psi^a_J =  \Phi^a_J$ on the space $\I_X \ss J^{(n)} M$ of $X$-invariant functions, identified by the vanishing of $D_J Q^a$ for all the multiindices $J$, $0 \le |J| \le (n-1)$. This also implies that the space $\I_X$ is invariant under $Y$, and $\mu$-symmetries (see below) can be used for symmetry reduction in the same way as standard symmetries. See \cite{GaMo} for details. \EOR

\medskip

Similarly to what happens for standard prolongations (and symmetries), discussing $\mu$-prolongations is simpler when we consider vector fields in evolutionary form; we recall that for $X$ given by (1.3), the evolutionary representative is $X_Q = Q^a (\pa / \pa u^a)$, with $Q$ as given above. Its $\mu$-prolongation $Y_Q$ is given by (1.4) with $\Psi^a_J = (\nabla_J )^a_b Q^b$; here we have written $\nabla_J := (\nabla_1)^{j_1} \cdots (\nabla_p)^{j_p}$, which is legitimate in view of (1.7$''$).

In the following, we will only deal with vector fields in evolutionary form. Note that in geometrical terms an evolutionary vector field is a (generalized) vector field on $(M,\pi,B)$ which is vertical for the fibration $\pi$.

The (standard or $\mu$) $k$-th prolongation of such a vector field will be a vector field on $(J^{(k)} M , \pi_k , B)$ which is vertical for the fibration $\pi_k$, and this for any $k$. We will denote by $\V_k$ the set of vertical vector fields on $(J^{(k)} M , \pi_k , B)$.

\subsection{Symmetries and $\mu$-symmetries}

The vector field $X$ is a {\bf $\mu$-symmetry} (with a given $\mu$) of the system of $r \ge 1$ differential equations $\De := (\De_1 , ... , \De_r )$ of order $n$ if and only if its $\mu$-prolongation $Y$ in the jet space $J^{(n)} M$ is tangent to the solution manifold $S_\De \ss J^{(n)} M$, $Y : S_\De \to \T S_\De$. Equivalently (under standard nondegeneracy assumptions), $X$ is a $\mu$-symmetry of $\De$ if there is a smooth matrix function $R : J^{(n)} M \to {\rm Mat}(r) $ such that $Y(\De_\a ) = R_\a^\b \, \De_\b$.

If $Y$ is tangent to all level manifolds for $\De$, i.e. if in the previous notation the function $R$ is identically zero, we say that $Y$ is a {\it strong} $\mu$-symmetry for $\De$.

\medskip\noindent
{\bf Remark 2.} Note that in order to consider $\mu$-symmetries of $\De$, it suffices that the compatibility condition (1.7) be verified on  $S_\De$, not necessarily in the whole jet space; we would then have ``internal $\mu$-symmetries'', in analogy with standard internal symmetries \cite{Kam}. This point will be relevant in section 4 below. \EOR

\medskip\noindent
{\bf Remark 3.} Note also that, as mentioned above, if $\La_i = \La_i (x,u^{(1)})$, then the $\mu$-prolongation of a Lie-point vector field in $M$ is a proper (rather than generalized) vector field in each jet space $J^{(k)} M$. \EOR

\section{The relation between $\mu$-prolongations and \\ standard prolongations}

In this section we show that $\mu$-prolonged vector fields are
locally related to vector fields which are standard prolongations. This will entail an equivalence between $\mu$-symmetries and (local or nonlocal, see sect.4) standard symmetries for a given equation.

We will denote by $G$ the group $GL(q,\R)$ of $q$-dimensional non-singular real matrices, and by $\G$ its Lie algebra. We denote by $\Ga_n$ the space of smooth maps $\ga: J^{(n)} M \to G$. The space $\Ga_n$ has a natural structure of Lie group.\footnote{In physical language, it is a {\it gauge group} modelled on $G$ \cite{Ish,Nak}. To avoid any confusion, note that usually -- that is, in Yang-Mills theories -- one considers gauge transformations depending on the base space variables only, i.e. $\ga : B \to G$; our setting is thus considerably more general.}

When $F$ is $q$-dimensional there is a natural action of the group $G=GL(q)$ on $T_f F$ for any point $f \in F$. As $(M,\pi,B)$ is a vector bundle, the jet bundle $(J^{(n)} M , \pi_n , B)$ is also a vector bundle.

The group $G$ acts naturally on the fiber $F = \R^q$; it also acts on the tangent space $\T_p F^{(n)}$ to the fibers $F^{(n)} = \pi_n^{-1} x$ of $J^{(n)} M$ at any point $p = (x,u^{(n)} ) \in \pi_n^{-1} x$. This is a vector space whose dimension $d=s q$ is an integer multiple of $q$. We will consider the $G$ action on $\T_p F^{(n)}$ via the $d$-dimensional representation which is the direct sum of $s$ copies of the defining one, the invariant subspaces spanned by $u^a_K$ for $a=1,..,q$ and a given multiindex $K$. We refer to this action as the {\it jet representation} of $G$.

If the jet representation of $G$ is given by the $d$-dimensional matrices $T : G \to {\rm Mat} (d)$, $T : g \mapsto T_g$, the function $\ga \in \Ga_n$ is represented by the matrix function $T \circ \ga : z \mapsto T_{\ga (z)} $ (with $z \in J^{(n)} M$). In the following we will use, with a standard abuse of notation, the same symbol $\gamma$ for the map $\ga : M \to G$ and for its $q\times q$ matrix representation. Thus $T_\ga = \ga \oplus ... \oplus \ga$.

A map $\ga \in \Ga_n$ acts naturally in $\V_n$ by the jet action. In fact, rewriting now a generic element $Y \in \V_n$ as 
$$ Y \ = \ \Psi^a_J \, (\pa / \pa u^a_J) \ , $$ 
$\ga$ acts on $Y$ to give $$ \ga \cdot Y \ := \ \[ \ga^a_b \ \Psi^b_J \] \ (\pa / \pa u^a_J) \ . \eqno(2.1) $$ 
This action obviously maps $\V_n$ into itself.

As mentioned above, we will prove that $\mu$-prolongations and ordinary ones are related by an action of the (gauge) group $\Ga_n$ on vertical vector fields. In order to do this, we need the following notions, which are obviously equivalence relations, due to the group property of $\Ga_n$.

\medskip\noindent 
{\bf Definition 1.} Given two vector fields $Y$ and $W$ in $\V_n$, we say that they are {\bf $G$-equivalent} if there exists a function $\ga \in \Ga_n$ such that $Y = \ga \cdot W$, globally in $J^{(n)} M$. We say that $Y$ and $W$ are {\bf locally $G$-equivalent} if for any $z \in J^{(n)}(M)$ there exists a neighborhood $A_z$ of $z$ and a local function $\ga_z : A_z \to G$ such that $W = \ga_z \cdot Y$ in $A_z$. \EOD 

\medskip

We will consider, together with the function $\ga \in \Ga_n$, the function $\ga^{-1} \in \Ga_n$. We stress that $\ga^{-1}$ is not the inverse of the function $\ga$, but a function which at each point $(x,u^{(n)})$ provides the element of $G$ inverse to the element $\ga (x,u^{(n)})$.

\subsection{Compatibility condition}

We start by identifying the geometrical meaning of the compatibility condition (1.7) in lemma 1. We will then recall proposition 1 (which we quote from \cite{Sha}, see chapter 3, theorem 6.1 there). Proposition 2 is given in \cite{Mar}.

\medskip\noindent 
{\bf Lemma 1.} {\it The compatibility condition (1.7) is the expression in coordinates of the horizontal Maurer-Cartan equation 
$$ D \mu \ + \ (1/2) \,  [ \mu , \mu ] \ = \ 0 \ ; \eqno(2.2') $$ 
equivalently, it states that the Maurer-Cartan equation is satisfied up to a form $\rho = - (\pa \La_j / \pa u^a_K ) \d x^j \vth^a_K$ in the Cartan ideal ${\cal J} (\Theta )$ generated by $\Theta$, 
$$ \d \mu \ + \ (1/2) \, [ \mu , \mu ] \ = \ \rho \, \in \, {\cal J} (\Theta ) \ . \eqno(2.2'') $$}

\medskip\noindent 
{\bf Proof.} The ``proof" amounts to recalling some definitions for $\G$-valued forms \cite{Sha,Str}. Let $\{ e_i \}$ be a basis in $\G$, so that any $\G$-valued form $\om$ (of any degree $k$) is written uniquely as $\om = e_i \otimes \om^i$, with $\om^i$ a $k$-form; one usually writes simply $\om = e_i \om^i$, omitting the tensor product symbol, for short.  Given two $\G$-valued forms $\a = e_i \otimes \a^i$ and $\b = \b^j \otimes e_j$, we define as usual \cite{Sha,Str} 
$$ [ \a , \b ] \ := \ [ e_i , e_j ] \, \otimes \, (\a^i \w \b^j ) \ . $$
Denote by $D$ the total exterior derivative operator; for a horizontal form $\a = A_\s \d x^\s$ (this is a shorthand for $\a = A_{\s_1 ... \s_k} \d x^{\s_1} \w ... \w \d x^{\s_k}$) its action is given by $ D \a = (D_i A_\s) \d x^i \w \d x^\s$. Using these definitions, we have indeed 
$$ \begin{array}{rl} 
D \mu \, + \, {1 \over 2} [ \mu , \mu ] \ =& \( D_i \La_j + {1 \over 2} \, [\La_i , \La_j ] \) \, \otimes \, (\d x^i \w \d x^j ) \ = \\ 
 =& \ {1 \over 2} \, \( D_i \La_j - D_j \La_i + [\La_i , \La_j ] \) \, \otimes \, ( \d x^i \w \d x^j ) \ , \end{array} $$ as claimed. 
Using (1.1) we have $D \mu = \d \mu - (\pa \La_j / \pa u^a_K ) \vth^a_K \w \d x^j$, hence the equivalent form of the statement. \EOP

\medskip\noindent 
{\bf Proposition 1.} {\it Let $E$ be a smooth manifold, $G$ a Lie group with Lie algebra $\G$, and $\mu$ a $\G$-valued one form on $E$ satisfying the structural (Maurer-Cartan) equation $$ \d \mu \ + \ {1 \over 2} \, [\mu, \mu] \ = \ 0 \ . $$ Then for each point $z \in E$ there are a neighborhood $A_z \sse E$ of $z$ and a local function $\ga_z : A_z \to G$ such that $\mu = \ga_z^{-1} \d \ga_z$ locally in $A_z$.}

\medskip\noindent 
{\bf Proposition 2.} {\it Let $(E,\pi,A)$ be a fiber bundle, $G$ a Lie group with Lie algebra $\G$, and $\mu$ a horizontal $\G$-valued one form on $E$ satisfying the horizontal Maurer-Cartan equation (2.2$'$). 
Then for each point $z \in E$  there are a neighborhood $A_z \sse E$ of $z$ and a local function $\ga_z : A_z \to G$ such that $\mu = \ga_z^{-1} D \ga_z$ locally in $A_z$.}

\medskip\noindent 
{\bf Definition 2.} Let $f : E \to G$ be a smooth map, $\om$ (respectively $\om_h$) the Maurer-Cartan (respectively, the horizontal Maurer-Cartan) form on $G$. Then the {\bf Darboux derivative} of $f$ is the $\G$-valued form $f^* (\om) = f^{-1} \d f \in \La^1 (E,\G )$ \cite{Mal,Sha}. The {\bf horizontal Darboux derivative} of $f$ is the horizontal $\G$-valued form $f^* (\om_h) = f^{-1} D f \in \La^1 (E,\G )$. \EOD

\medskip\noindent 
{\bf Remark 4.} If we denote by $\om$ the Maurer-Cartan form on $G$, proposition 1 guarantees that any $\G$-valued one-form $\mu$ satisfying the structural equation is (locally) the Darboux derivative of some (local) function $\ga$. \EOR

\subsection{Prolongations}

We proceed now, equipped with the previous results, to prove lemma 2 and theorem 1 (and 2) which establish the (local) $G$-equivalence of standard and $\mu$-prolongations.

\medskip\noindent 
{\bf Lemma 2.} {\it Let $\G$ be the Lie algebra of the group $G$, and let $\mu$ be a $\G$-valued horizontal one form on $(J^{(n)} M,\pi_n ,B)$, written in local coordinates as $\mu = \La_i \d x^i $. Then $\mu$ satisfies the compatibility condition (1.7) if and only if for any $z \in J^{(n)} M$ there are a neighborhood $A_z \sse J^{(n)} M$ and a function $\ga_z : A_z \to G$  such that 
$$ \La_i \ := \ \ga_z^{-1} (D_i \ga_z) \ . \eqno(2.3) $$}

\medskip\noindent 
{\bf Proof.} With the choice (2.3), we have $D_i \La_j = (D_i \ga_z^{-1} )  (D_j \ga_z ) + \ga_z^{-1}  (D_i D_j \ga_z )$; moreover, $\La_i \La_j = \ga_z^{-1} (D_i \ga_z ) \ga_z^{-1} (D_j \ga_z) = - (D_i  \ga_z^{-1} ) (D_j \ga_z )$. These show at once that (1.7) is satisfied when $\La_i$ are given by (2.3).
As for the converse, i.e. that if (1.7) is satisfied then locally $\mu$ is written as $\mu = \La_i \d x^i$ with $\La_i$ as in (2.3), this follows from  lemma 1 and proposition 2. \EOP

\medskip\noindent 
{\bf Remark 5.} If the function $\ga$ is given, the horizontal form $\mu = \La_i \d x^i$ is uniquely determined by (2.3). On the other hand, if a horizontal $\mu$ -- satisfying (1.7) -- is given, the function $\ga$ is not uniquely determined by (2.3), even locally in $A_z$. Indeed, if $\ga_1$ satisfies (2.3), we can always consider $\ga_2 = h \ga_1$, with $h$ any constant element of $G$, which still satisfies (2.3). \EOR

\medskip\noindent 
{\bf Theorem 1.} {\it Let $Y=\Psi^a_J (\pa / \pa u^a_J) \in \V_n$ be the $\mu$-prolongation of the (possibly generalized) vector field $X = \Psi^a_0 (\pa /  \pa u^a ) \in \V_0$. Suppose that $\mu=  \ga^{-1} (D \ga) = \ga^{-1} (D_i \ga) \d x^i$ for a smooth function $\ga \in \Ga_n$. Then the vector field $W = \ga \cdot Y \in \V_n$ is the standard prolongation of the (possibly generalized) vector field $\XT = \ga \cdot X \in \V_0$.
Conversely, let $W := \XT^{(n)} = \PT^a_J (\pa / \pa u^a_J)$ be the standard prolongation of the evolutionary vector field $\XT = \PT^a_0 (\pa / \pa u^a) \in \V_0$; let $\ga \in \Ga_n$ be a smooth function. Then the vector field $Y \in \V_n$ defined by $Y = \ga^{-1} \cdot W$ is the $\mu$-prolongation of the (possibly generalized) vector field $X = \ga^{-1} \cdot \XT$, with $\mu = \ga^{-1} (D \ga)$.}

\medskip \noindent 
{\bf Proof.} Saying that $W = \XT^{(n)}$ is a standard prolongation means that $$ \PT^a_{J,i} \ = \ D_i ( \PT^a_J ) \ . $$
On the other hand, $W = \ga \cdot Y$, hence $\PT^a_K = \ga^a_b \Psi^b_K$ for any multiindex $K$, see (2.1). In vector notation,
$$ \PT^a_{J,i} := \ \ga^a_b \Psi^b_{J,i} \ = \ 
D_i ( \ga^a_b \Psi^b_J ) \ = \ \ga^a_b (D_i \Psi^b_J) \ + \ \ga^a_b [(\ga^{-1})^b_c (D_i \ga^c_m)] \, \Psi^m_J \ . $$ 
As the matrices $\ga (x,u^{(n)} )$ are invertible, this shows that $$ \Psi^a_{J,i} \ = \ \[ \de^a_b D_i \, + \, (\ga^{-1})^a_k (D_i \ga^k_b) \] \Psi^b_J \ , $$ 
which is just the $\mu$-prolongation formula with the identification $ \La_i := \ga^{-1} (D_i \ga) $, as claimed in the statement.
In order to prove the converse, it suffices to perform the computation the other way round, i.e. starting with $\Psi^a_{J,i} = \nabla_i ( \Psi^a_J )$. \EOP 

\medskip

In the theorem above, $\mu$ is known to be in the form $\mu = \ga^{-1} (D \ga )$; we will now consider the general case.

\medskip\noindent 
{\bf Theorem 2.} {\it Let $Y \in \V_n$ be the $\mu$-prolongation of the vector field $X \in \V_0$, with $\mu$ satisfying (2.2). Then for any $z = (x , u^{(n)} ) \in J^{(n)} M$ there are a neighborhood $A_z \sse J^{(n)} M$ and a local map $\ga_z : A_z \to G$ such that, locally in $A_z$:
\par\noindent {\tt (i)} the form $\mu \in \La^1 (J^{(n)} M, \G )$ is given by $\mu = \ga_z^{-1} (D \ga_z)$; 
\par\noindent {\tt (ii)} $Y$ is $G$-equivalent to the vector field $W := \ga_z \cdot Y$, which is the standard prolongation of the vector field $\XT := \ga_z \cdot X \in \V_0$.}

\medskip\noindent 
{\bf Proof.} This merely states that theorem 1, which deals with the case where $\mu = \ga^{-1} (D \ga)$ for some map $\ga \in \Ga_n$, actually holds {\it locally} for all $\mu$ satisfying (2.2), or equivalently (1.7). In fact, lemma 2 guarantees that any such $\mu$ can be written, locally in $A_z$, in the form $\mu = \ga_z^{-1} (D \ga_z)$. \EOP 
\medskip

We stress that proposition 1, and hence lemma 2 and theorem 2, make only local claims. Let us now consider the global setting.

\medskip\noindent {\bf Theorem 3.} {\it Let $\mu = \La_i \d x^i$ be a $\G$-valued horizontal one form on $J^{(n)} M$ satisfying (1.7), equivalently (2.2). Let $Y=\Psi^a_J (\pa / \pa u^a_J) \in \V_n$ be the $\mu$-prolongation of the (possibly generalized) vector field $X = \Psi^a_0 (\pa / \pa u^a ) \in \V_0$. Then the following are equivalent: 
\par\noindent {\tt (i)} There exists a vector field $W \in \V_n$ which is $G$-equivalent to $Y$ and which is the standard prolongation of a vector field $\XT \in \V_0$. The vector fields $W$ and $\XT$ are given by $W = \ga \cdot Y$ and $\XT = \ga \cdot X$, with $\La_i = \ga^{-1} (D_i \ga)$. 
\par\noindent {\tt (ii)} $\mu$ is the horizontal Darboux derivative of a $\ga \in \Ga_n$, $\mu = \ga^{-1} D (\ga)$.}

\medskip\noindent {\bf Proof.} This follows at once from theorem 1 and lemma 2. \EOP 
\medskip

We denote by $\M$ the set of $\G$-valued horizontal one-forms satisfying the horizontal Maurer-Cartan equation and by $\D$ the set of one-forms which are the Darboux derivative of maps $\ga \in \Ga_n$; by lemma 2, $\D \sse \M$. We define $\H := \M / \D$; this amounts to factorization with respect to gauge equivalence. Note that $\H$ depends on the topology (in particular, the first homotopy group) of $J^{(n)} M$; when the fibers of $M$ and hence of $J^{(n)} M$ are contractible -- as in our case, where $M$ is a vector bundle -- this is just the topology of the base space $B$ (see examples 2 and 3 below in this respect).

\medskip\noindent {\bf Corollary 1.} {\it If $\H = 0$, then any $\mu$-prolonged vector field, for any $\mu \in \M$, is globally $G$-equivalent to a standard prolongation. In particular, this holds if $J^{(n)} M$ is contractible.}

\medskip\noindent {\bf Proof.} Obvious. \EOP 

\medskip

The previous results, in particular theorem 3 and its corollary, could also be obtained from the global version of proposition 1, see again \cite{Sha} (in particular chapter 3, theorem 7.14). This states that $\mu$ is the Darboux derivative of a map $\ga : J^{(n)} M \to G$ if and only if the monodromy representation $\Phi_\mu : \pi_1 (J^{(n)} M, z) \to G$ is trivial (for any point $z \in J^{(n)} M$) and $\mu$ satisfies the Maurer-Cartan equation.

\medskip\noindent 
{\bf Remark 6.} When we consider the scalar case $q=1$, the group $G$ reduces to the multiplicative group $\R_+$, hence the $\La_i$ reduce to nonzero smooth functions $\la_i : J^{(n)} M \to \R$; two vertical vector fields $Y$ and $W$ in $\V_n$ are $G$-equivalent if and only if they are collinear, i.e. if there is a (scalar) nowhere zero function $\ga : J^{(n)} M \to \R$ such that $W = \ga Y$. Note that in this case $G$ is an abelian group, and (1.7) reduces to $D_i \la_j = D_j \la_i$, where the $\la_i$ are real functions; this means that locally there is a ``potential'' $V$ such that $\la_i = D_i V = D_i(\ga)/\ga$. \EOR

\medskip\noindent 
{\bf Remark 7.} If we consider ODEs rather than PDEs, we have $B=\R$ and $p = 1$. In this case condition (1.7) is automatically satisfied, and does not impose any restriction on the $\La_i$. \EOR

\medskip\noindent 
{\bf Remark 8.} The work \cite{Sac} considered scalar ODEs; it was shown there that vector fields $Y$ on $J^{(n)} M$ which are the $\la$-prolongation of a vector field $X$ in $M$ (with standard prolongation $W = X^{(n)}$) can be identified by the property that the characteristics of $W$ and of $Y$ coincide. In the frame of remark 6 above, having a one-dimensional fiber ($q=1$) makes that $GL(q)$ reduces to the group of nowhere vanishing real-valued functions $\la (x,u^{(n)} )$, and vertical vector fields are just scaled by $\la$, hence preserving the characteristics. In the general case, our result says that the characteristics are acted upon by the group $G$, in a covariant way: that is, acting on the vectors $u_K = (u^1_K , ... , u^q_K)$ (for any multiindex $K$, $|K| \ge 0$) in the same way. This is precisely the requirement that $G$ acts by the jet action, see above. \EOR

\medskip\noindent 
{\bf Remark 9.} In the language of gauge theories, $\ga = \ga (x)$ and the $\La_i$ satisfying (1.7) identify a flat connection, i.e. a zero strength Yang-Mills field. Thus proposition 1 is a generalization (see footnote 1) of the statement that any such field reduces locally to a pure gauge field; see e.g. theorem 6.4 in \cite{Ish}. Similarly, lemma 2 above identifies conditions to have a pure gauge field (in our extended sense) at the global level as well. \EOR

\subsection{Symmetries}

We now pass to discuss symmetries. As pointed out in remark 2, it would actually suffice that $\mu$ satisfy (1.7) on $S_\De \ss J^{(n)} M$. On the other hand, the distinction between local and global (in $S_\De$) $G$-equivalence still applies. The result obtained earlier on and these observations lead to the following result.

\medskip\noindent
{\bf Theorem 4.} {\it Let $\De$ be an equation or system of equations in $J^{(n)} M$. Let $\mu = \La_i \d x^i$ be a horizontal $\G$-valued one-form on $J^{(n)} M$, such that (1.7) are verified on a manifold $C_\mu \sse J^{(n)} M$, with $S_\De \sse C_\mu$. Let the (possibly generalized) vector field $X = \Psi^a_0 (\pa / \pa u^a ) \in \V_0$ be a $\mu$-symmetry for $\De$. Then the following are equivalent: \par\noindent {\tt (i)} There exists a (possibly, generalized and/or nonlocal) vector field $\XT \in \V_0$ which is $G$-equivalent to $X$ and which is a standard symmetry of $\De$; this is given by $\XT = \ga \cdot X$, where $\mu = \ga^{-1} (D \ga)$ on $C_\mu$ and hence in particular on $S_\De$. 
\par\noindent {\tt (ii)} The form $\mu$ is the horizontal Darboux derivative of $\ga : C_\mu \to G$.}

\medskip\noindent 
{\bf Proof.} For $C_\mu = J^{(n)} M$, this follows at once from theorem 3 and the definitions of standard and $\mu$-symmetry of $\De$. In the general case, we just have to restrict all forms, operators (and considerations) to the manifold $S_\De \sse C_\mu \sse J^{(n)} M$ of interest when discussing symmetries of $\De$. \EOP 

\medskip

Note that the possibility of having a nonlocal vector field, mentioned in this statement, is present only in the case where $C_\mu \not= J^{(n)} M$, as discussed in detail in section 4.

\section{Conditional and partial $\mu$-symmetries}

In this section we define conditional and partial $\mu$-symmetries, and discuss the relation of these with ``standard'' conditional and partial symmetries; the relevant definitions for the latter will also be recalled. We will work at the local level only; discussion of global problems would go along the same lines of section 2.2 (in particular, our results hold also globally if $\mu$ is the Darboux derivative of a function $\ga \in \Ga_n$, see theorems 3 and 4 above) and is left to the interested reader.

\subsection{Conditional symmetries}

Let us consider first of all conditional symmetries: they are most frequently considered in the scalar case $q=1$, but we can deal with the general case $q>1$.

\medskip\noindent 
{\bf Definition 3.} The vector field $X$ is a (standard) {\bf conditional symmetry} of the differential system $\De$ if there are solutions $u^a = f^a(x)$ to $\De$ which are invariant under $X$, i.e., such that $X : \sigma_f \to \T_{\sigma_f}$, where $\sigma_f$ is the associated section in $M$. If $X$ is not a symmetry of $\De$, we say it is a {\it proper} conditional symmetry. \EOD 
\medskip

A particularly convenient way to look at conditional symmetries of a differential system of $r$ equations $\De := \De_\alpha (x,u^{(n)} ) = 0$ is the following (cf. \cite{LeWi,Win}).

We consider the system made of $\De$ and of the equations expressing the $X$-invariance of the solution  $u^a = f^a(x)$. The latter are simply $Q^a =  0$, with $Q^a$ the characteristic of the vector field $X$, see above. That is, we are considering the system of $r +q$ equations
$$ \cases{ \De_\alpha (x,u^{(n)}) \ = \ 0 \ , & \cr 
\phi^a - u^a_i \xi^i \ = \ 0 \ . & }  \eqno(3.1) $$
Any solution to (3.1) is $X$-invariant and a solution to $\De$, and conversely all the $X$-invariant solutions to $\De$ are solutions to the system (3.1). Then, the conditional symmetries for $\De$ are simply those vector fields $X$ for which (3.1) admits solutions. Note that $X$ turns out to be a symmetry of the system (3.1) (see however \cite{OR,PS,Sac} and also \cite{CiK} for a careful discussion on the notion of conditional symmetries).

\medskip

If $X$ is a symmetry of $\De$, it can  happen -- as well known, also in the case $q=1$ (see \cite{OR})   -- that $\De$ does not admit any $X$-invariant solution (i.e., that the system (3.1) does not admit solutions, for a given $X$), even when $X$ is an exact symmetry of $\De$. In this sense, standard (exact) symmetries are not necessarily conditional symmetries as well.

In the computation of conditional symmetries one often, to obtain simpler formulas, normalizes to one a nonzero coefficient in the vector field $X$ (see, e.g., \cite{Win}). This procedure can be seen as a special case (see remark 6) of the following simple lemma.

\medskip\noindent 
{\bf Lemma 3.} {\it Let $X_0$ be a conditional symmetry of $\De$, in evolutionary form; let $\ga \in \Ga_n$. Then $X  := \ga \cdot X_0$ is also a conditional symmetry of $\De$.}

\medskip\noindent 
{\bf Proof.} The prolongations of $X$ and of $X_0$ differ by terms which vanish on the set where $D_J Q^a = 0$ for all $|J|=0,...,n-1$. \EOP

\subsection{Conditional $\mu$-symmetries}

A natural definition of conditional $\mu$-symmetries would be the one below.

\medskip\noindent
{\bf Definition 4.} The vector field $X$ is a {\bf conditional $\mu$-symmetry} of the differential system $\De$ if it is a $\mu$-symmetry of the system (3.1). If $X$ is not a $\mu$-symmetry of $\De$, we say that it is a {\it proper} conditional $\mu$-symmetry. \EOD

\medskip\noindent 
{\bf Remark 10.} As in the case of standard conditional symmetries, a $\mu$-symmetry $X$ of the system $\De$ could fail to be also a conditional $\mu$-symmetry (as $\De$ may do not admit solutions invariant under the $\mu$-prolongation of $X$);  an example with $q=1$ is $\De := e^x u_x + u_y- 1=0$, with $X = \pa_x +  e^{-x} \pa_y $ and $\mu = \d x$.  \EOR

\medskip 
One could prove the analogue of theorem 4, i.e. that for any conditional $\mu$-symmetry $X$ there is a $G$-equivalent vector field $\XT$ which is a standard conditional symmetry. However, as stated by lemma 3, standard  conditional symmetries already come in $G$-equivalent families, so that  considering conditional $\mu$-symmetries does not give anything new with  respect to standard conditional symmetries:

\medskip\noindent 
{\bf Corollary 2.} {\it Let $\De$ be a differential system; the vector field $X$ is a conditional $\mu$-symmetry for $\De$ if and only if it is a standard conditional symmetry for $\De$.}

\medskip\noindent 
{\bf Proof.} By definition, a vector field $X$ is a conditional $\mu$-symmetry (resp. a standard conditional symmetry) if it is a $\mu$-symmetry (resp. a standard symmetry) of the system (3.1). Now, the $\mu$-prolongations and the standard prolongations coincide on the invariant set $\I_X\sse J^{(n)}M$ identified by the conditions $Q^a=0$ and their differential consequences, but in looking for conditional symmetries one restricts precisely to this set. \EOP

\medskip

We can also investigate the relation between standard conditional symmetries and ``full'' (rather than conditional) $\mu$-symmetries. It turns out that there is a correspondence  between $\mu$-symmetries and a special subset of (standard) conditional symmetries.

\medskip\noindent 
{\bf Corollary 3.} {\it Let $X$ be a full $\mu$-symmetry for a system of PDEs $\De$, and let $\De$ admit a $X$-invariant solution. Then $X$ is a (standard) conditional symmetry for $\De$. 
More precisely, $X$ is a full $\mu$-symmetry of $\De$, admitting  $X$-invariant solutions, if and only if: {\rm (i)} $X$ is a (standard) conditional symmetry of $\De$, and {\rm (ii)} there is $\ga \in \Ga_n$ such that $\ga \cdot X$ is an exact symmetry of $\De$.}

\medskip\noindent 
{\bf Proof.} According to theorem 4, if $X$ is a (full) $\mu$-symmetry of $\De$, there is a $G$-equivalent exact symmetry $\XT = \ga \cdot X$. Conversely, if $\XT$ is an exact symmetry in evolutionary form, then, for any $\be \in \Ga_n$, one has that $X = \be \cdot \XT$ is a $\mu$-symmetry, with $\mu= \be (D \be^{-1})$. The conclusion follows observing that if $\XT$ is an exact symmetry (admitting some invariant solution) then $\be\cdot\XT$ is in general no longer an exact symmetry but only a conditional symmetry. \EOP

\medskip\noindent
{\bf Remark 11.} It is well known that conditional symmetries of a given equation do not transform solutions of the given equation into other solutions; the same is true, as already remarked, for $\mu$-symmetries.  It should be stressed  that the latter have a more convenient property: indeed, under  the conditions of theorem 4, the presence of a $\mu$-symmetry $X$ guarantees that there exists a true $G$-equivalent symmetry $\XT$ which maps solutions into solutions. \EOR

\subsection{Partial symmetries}

Partial symmetries were introduced in \cite{CGpar}, and represent a generalization of conditional symmetries. The vector field $X$ is a (proper) {\bf partial symmetry} of the differential equation $\De$ if there is a (proper) subset ${\cal S}$ of solutions which is mapped into itself under $X$. In a sense, these interpolate between standard and conditional symmetries: if ${\cal S}$ is the set of all solutions to $\De$, we have a standard symmetry (in this case $X$ is a trivial partial symmetry), and if there is some solution which is invariant under $X$ (and hence in ${\cal S}$) we have a conditional symmetry. Note that we are {\it not} requiring there is any $X$-invariant solution. See \cite{CGpar} for details.

For the sake of simplicity, we restrict here our discussion to the case $q=1$; the extension to the general case is once again completely straightforward (example 7 below will consider the case $q=2$).

We can look at partial symmetries in a way similar to the Levi and Winternitz approach to conditional symmetries \cite{LeWi}. We apply the vector field $X^{(n)}$, where $n$ is the order of $\De$, to $\De$ and restrict the obtained expression to the solution manifold $\S_\De$; we call this $\De^{(1)}$. Note that $\De^{(1)}$ cannot vanish unless $X$ is a symmetry of $\De$. We repeat then the procedure, i.e. define $$ \De^{(k+1)} \ := \[ X^{(n)} (\De^{(k)} ) \]_{\S_k} \eqno(3.2) $$ where $\S_k$ is the intersection of the solution manifolds for $\De\equiv \De^{(0)}, \De^{(1)} ,..., \De^{(k)}$. We are interested in the case there is some finite $\ell$ such that $\De^{(\ell+1)} \equiv 0$; in this case we say that $X$ is a partial symmetry of order $\ell$. Note that the set ${\cal S}={\cal S}_\ell$ corresponds to solutions to the system $\{ \De^{(0)} , \De^{(1)} , ... , \De^{(\ell)} \}$, and that $X$ is a standard symmetry for this system.

\medskip\noindent 
{\bf Lemma 4.} {\it Let $\De$ be a differential equation of order $n$, and $X_0$ any vector field, on $M$; denote by $W:= X_0^{(n)}$ the standard $n$-th order prolongation of $X_0$. Consider a vector field $Y = \be \cdot W$ in $J^{(n)} M$, with $\be \in \Ga_n$, and write $\^\De^{(k)}$ for the restriction to $\^S_{k-1}$ of $(Y)^k [ \De ]$, with $\^S_k$ the intersection of the solution manifolds for $\De^{(0)}, \^\De^{(1)} ,..., \^\De^{(k)}$. Then $S_k = \^S_k$, and $\^\De^{(k+1)} \equiv \De^{(k+1)}$ on $\S_k$, for all $k \ge 0$.}

\medskip\noindent 
{\bf Proof.} The action of $Y$ on $\De$ will produce $\^\De^{(1)} := Y [\De] = \be  \De^{(1)}$. Further applying $Y$ we obtain $\^\De^{(2)} := Y[\^\De^{(1)}]=\be W[ \be \De^{(1)}] = \be^2 \De^{(2)} + [\be W(\be)] \De^{(1)}$. In this way it is easy to see that we obtain, for all $k$, 
$$ \^\De^{(k)} \ := \ \[ (Y)^k [\De] \]_{\^S_{k-1}} \ = \ \[ \be^k \De^{(k)} + \sum_{m=0}^{k-1} h_m \De^{(m)} \]_{\^S_{k-1}} \eqno(3.3) $$ where $h_m$ are smooth functions.
It is obvious from (3.3) that $\^\De^{(k+1)} = \De^{(k+1)}$ on $\S_k$, and actually that $\S_k = \^S_k$ for all $k$. \EOP

\subsection{Partial $\mu$-symmetries}

As for conditional symmetries, we can parallel the definition of (standard) partial symmetries and define partial $\mu$-symmetries. Roughly speaking, theorem 4 extends also to this case.

\medskip\noindent
{\bf Definition 5.} The vector field $X$ is a (proper) {\bf partial $\mu$-symmetry} of the differential equation $\De$ if there is a (proper) subset ${\cal S}$ of solutions $u = f(x)$ to $\De$, invariant under the $\mu$-prolongations $Y$ of $X$. \EOD

\medskip

This definition can be recast in a constructive way in this form [cf. (3.2)]:

\medskip\noindent 
{\bf Definition 5$'$}. Let $\De=0$ be a PDE, $X$ a vector field and $Y$ a $\mu$-prolongation of $X$. Assume that $X$ is not a $\mu$-symmetry for $\De$, i.e. $\De^{(1)}:=[Y(\De)]|_{\De=0}\not= 0$, but, defining $$\De^{(k+1)}:=[Y(\De^{(k)})|_{\S_k} \quad {\rm for}\quad k=0,1,\ldots,\ell \eqno(3.4)$$ with the same notations as above, assume that $\De^{(\ell+1)}\equiv 0$. We then say that  $X$ is a (proper) {\bf partial $\mu$-symmetry of order $\ell>1$}. \EOD

\medskip\noindent 
{\bf Corollary 4.} {\it Let $\De$ be a differential equation of order $n$, and $X$ a (proper) partial $\mu$-symmetry for $\De$; denote by $Y$ the $\mu$-prolongation of order $n$ of $X$. Then there is $\ga \in \Ga_n$ such that $\XT := \ga \cdot X$ is a (proper) standard partial symmetry of $\De$.}

\medskip\noindent {\bf Proof}. This is a consequence of theorem 4 and of lemma 4, with $X_0=\XT$ and $\be=\ga^{-1}$. \EOP

\bigskip 

Partial $\mu$-symmetries differ from full $\mu$-symmetries also for what concerns the reduction of PDEs, i.e. the problem of finding $X$-invariant solutions.  In the case of $\mu$-symmetries the reduction -- which, as already remarked,  is performed via the same  method as for standard symmetries \cite{GaMo} -- leads to a new equation, say $\De^{[0]}(z,w)=0$, which involves the $X$-invariant variables $z$ and $w$ determined by the invariance condition $Q=0$. In the case of partial $\mu$-symmetries, instead, the PDE is transformed into an equation of the form (cf. \cite{OR,PS,Sac}, see also \cite{CiK}) 
$$ R_1(s,z) \, \De^{[0]}_1 (z,w) \, + \, R_2 (s,z) \, \De_2^{[0]} (z,w) \, + \, \ldots \ = \ 0 \eqno(3.5) $$ with coefficients $R_a$ depending also on some non-invariant coordinate (denoted by $s$); now the invariant solutions of the PDE can  be obtained solving the {\it system} of equations $\De^{[0]}_a=0$. In the case of a PDE involving a single function $u$ depending on two independent variables only, this becomes a system of ODEs (resp. a single ODE when $X$ is a full $\mu$-symmetry).

\medskip\noindent {\bf Remark 12.} It follows immediately from this argument and from lemma 4 that if $X$ is a partial $\mu$-symmetry of order $\ell$, and $\S=\S_\ell$ is the corresponding subset of solutions to $\De$, then there exists  a standard partial symmetry $\XT$ of the same order $\ell$, and with the same set $\S$ of solutions being globally invariant under $\XT$ as well. \EOR

\section{The relation between local $\mu$-symmetries and \\ 
nonlocal standard symmetries}

In this section we want to point out relations between $\mu$-symmetries and ordinary symmetries by using results of section 2 above. We will show that there is (locally) a one to one correspondence between local $\mu$-symmetries and  nonlocal standard symmetries of exponential form.

We recall that in the definition of a $\mu$-symmetry for  a system of differential equations $\Delta=(\Delta_1, \ldots \Delta_r)$, in order to apply the prolonged vector field on the solution manifold of the system, we only need that compatibility conditions (1.7) hold on the solution manifold $S_{\Delta}$ and not necessarily for all points in $J^{(n)}M$. In this case we cannot guarantee that for each point $z \in J^{(n)}(M)$ there are a neighborhood $A_z \subset J^{(n)} (M)$ of $z$ and a function $\ga : A_z \to G$ such that $\mu = \ga^{ -1 } (D_i \ga) \d x^i$ on $A_z$. In other words, if (1.7) holds only on the solution manifold $S_\De$, or however on a subset $C_\mu \not= J^{(n)} M$ of the jet space (with $S_\De \sse C_\mu$), we cannot have $G$-equivalence of $\mu$-symmetries with ordinary symmetries in the usual sense.

What we can prove is that there is a $G$-equivalence between $\mu$-symmetries and nonlocal symmetries of exponential form. We discuss here this equivalence only in the scalar case (i.e for the case $q=1$ of a single PDE), for ease of notation; the case of systems of PDEs is analogous.

Let $(M, \pi, B)$ a vector bundle with $1$-dimensional fiber $F=\pi^{-1}(x)=\R$ and $B=\R^p$. We call {\bf non-local exponential vector field} $X$ the formal vector field on $M$ given by
$$ \XT \ = \ e^{\int P_i(x,  u^{(n)}) \d x^i } \ X \ , \eqno (4.1) $$ 
where $X = \xi^i (\pa / \pa x^i) + \varphi (\pa / \pa u)$ is a (possibly generalized) vector field on $M$, and $P : J^{(n)} M \to \R^p$ is a vector function defined on $J^{(n)}M$, satisfying 
$$ D_k \, \[\int P_i (x, u^{(n)}) \, dx^i \] \ = \ P_k (x,  u^{(n)}) \ . \eqno(4.2) $$ 
The integral $\int P_i(x, u^{(n)}) \, dx^i$ is in general a formal expression. This generalizes the standard definition of nonlocal exponential vector field for ODEs \cite{Olv1}.

\medskip\noindent 
{\bf Definition 6.} 
Given a scalar PDE $\De$ we say that the exponential vector field (4.1) is a {\bf  nonlocal  exponential symmetry} of $\De$ if $D_i P_j = D_j P_i$ on the solution manifold $ S_{\De}$, and $[X^{(n)}(\De)]_{S_\De} = 0 $. \EOD 

\medskip

Now we can state our result relating $\mu$-symmetries and nonlocal exponential symmetry of $\Delta$. This extends theorem 5.1 in \cite{MuR1}.

\medskip\noindent 
{\bf Theorem 5.} {\it Let $X$ be a vector field on $M$, and $\mu = P_i (x,u^{(n)}) \d x^i$ a horizontal form, such that $D_i P_j = D_j P_i$ in $C_\mu \sse J^{(n)} M$. Let $\De$ be a PDE of order $n$ on $M$, such that $S_\De \sse C_\mu$. Let $X$ be a $\mu$-symmetry of $\De$. Then the nonlocal exponential vector field $\XT = \exp [\int P_i (x,u^{(n)} ) \d x^i ] \, X$ is a nonlocal exponential symmetry of $\De$.  Conversely, if $\XT$ as above is a nonlocal exponential symmetry of $\De$ with $X$ a vector field in $M$, then $X$ is a $\mu$-symmetry for $\De$.}

\medskip\noindent  {\bf Proof.} 
By using the results of section 2, it suffices to show that the ordinary prolongation $W=\XT^{(n)}$ of $\XT$ and the $\mu$-prolongation $Y$ of $X$ (with $\mu= P_i \d x^i$) are $G$-equivalent through the function  $$ \ga \ := \ \exp \[ \int P_i (x, u^{(n)}) \, \d x^i \] \ . $$ We know from Theorem 1 that $\mu$-prolongations (with $\mu = P_i \d x^i$) are $G$-equi\-va\-lent to ordinary prolongations by a function $\ga \in \Ga_n$ satisfying $\ga^{-1} D_i \ga = P_i$; the general (formal) solution of the previous equation for the unknown function $\ga$ is given by $\ga = \exp[\int P_i (x, u^{(n)}) \d x^i ]$. \EOP

\medskip\noindent  
{\bf Remark 13.} Determining the standard nonlocal symmetries of a differential equation is in general a very difficult problem. On the other hand, determination of local $\mu$-symmetries goes essentially through the same procedure as determination of standard (Lie-point or generalized) symmetries, i.e. we have a set of determining equations; see \cite{GaMo}. \EOR

\medskip\noindent {\bf Remark 14.} If $\mu$ is written globally (on $S_\De$) as $\mu = \ga^{-1} D \ga$ with $\ga : S_\De \to G$, see section 2, we have $\XT = \ga \cdot X$ on $S_\De$. \EOR

\section{Discussion}

In this short section we present some further general remarks on our approach.

\medskip\noindent
(1) First of all we note that although our motivation resided in getting a sound understanding of $\lambda$ and $\mu$-symmetries, and their relation with standard ones, which we believe is reached here, our result can also be seen the other way round. That is, we have actually obtained a description of how {\it gauge symmetries}, so important and pervading in modern Physics based on variational principles (e.g., Yang-Mills theories), enter in the theory of symmetries of -- in general, non-variational -- general differential equations.

\medskip\noindent
(2) We recall that it was shown in \cite{GaMo} that the determination of $\mu$-symmetries goes through the solution of determining equations pretty much as for standard symmetries. Note that even if one is interested only in standard symmetries, obtaining $\mu$-symmetries allows then to determine standard symmetries which are gauge-equivalent to these. 

\medskip\noindent
(3) The key role in our discussion -- and so in gauge symmetries of (non variational) differential equations -- is played by horizontal forms satisfying the horizontal Maurer-Cartan condition. These are also known as zero-curvature representations, and considerable work has been devoted to their study also from quite differents points of view. Here we just refer the reader to \cite{Mar} and to the recent volume \cite{AAM},  with the references in the works included therein, for this.

\medskip\noindent
(4) The examples we provide (see next section) are mainly intended to show that $\mu$-symmetries are -- in a suitable neighbourhood --  reducible to ordinary symmetries, maybe nonlocal. This corresponds to our main motivation in enterprising this work, see above, but could seem a bit disappointing to the reader seeking ``new'' symmetries. We would thus like to stress that one could as well manifacture examples of $\mu$-symmetries which are {\it not}  reducible to ordinary ones, e.g. by making sure that a nontrivial topology (in particular, horizontal cohomology) is present, as in examples 2 and 3. 

\medskip\noindent
(5) In the analysis of Yang-Mills theories, one is naturally led to study gauge fields in terms of the topology of a certain principal fiber bundle. The same applies here; more precisely, $\mu$-symmetries are locally equivalent to standard ones, and the way in which different local standard symmetries -- hence neighbourhoods -- are patched together to make a $\mu$-symmetry depends on the topology of the associated principal bundle $(P,\pi_G,J^{(n)} M)$ of fiber $G$ over the relevant jet space; see section 2.

\medskip\noindent
(6) For what concerns the application to nonlocal symmetries, not only the approach based on $\mu$-symmetries appears to be somewhat simpler (see also remark 13 above), but it may even happen that generalized vector fields with trivial characteristic vector provide nontrivial $\mu$-symmetries, and hence nontrivial nonlocal symmetries. This is shown in example 9, where we recover in this way some of the nonlocal symmetries of the Calogero-Degasperis-Ibragimov-Shabat equation studied by Sergyeyev and Sanders \cite{SeS}.

\medskip\noindent
(7) Finally, we would like to point out two possible directions of further developement. One of these concerns potential symmetries \cite{BK}, as suggested in example 10 below. A different direction (we thank one of the referees for suggesting this) concerns the relation of $\mu$-symmetries with pseudosymmetries \cite{Sok}; the latter are related to factorization with respect to symmetries rather than to invariant solutions.

\section{Examples}

In this section we consider examples illustrating our discussion of previous sections, both for the scalar and the vector case, considering full $\mu$-symmetries and partial $\mu$-symmetries as well.

Example 1 shows a global $G$-equivalence between a $\mu$-symmetry and a standard one, while in examples 2 and 3 the $G$-equivalence is only local. In examples 4 and 5 we deal with partial $\mu$-symmetries for scalar equations (KdV and Boussinesq, respectively), while in examples 6 and 7 we deal with full and partial $\mu$-symmetries of a system of PDEs. Finally, examples 8 and 9 illustrate the correspondence between local $\mu$-symmetries and nonlocal standard symmetries of exponential type, and example 10 recovers a well known case of potential symmetry in terms of $\mu$-symmetries.

\subsubsection*{Example 1.}

In example 1 (sect.6.2) of \cite{GaMo}, one considered the vector field $X = x \pa_x + 2 t \pa_t + u \pa_u$ and the form $\mu = \la \d x$, where $\la$ is a real constant. 
In this case, $\ga = e^{\la x}$; hence $$ \XT \ = \ e^{\la x} \ \( x \pa_x + 2 t \pa_t + u \pa_u \) \ . $$

It is easy to check that the functions on $J^{(2)} M$ which have been shown in \cite{GaMo} to be invariant under the $\mu$-prolongation of $X$, are in fact also invariant under the standard prolongation of $\XT$.

\subsubsection*{Example  2.}

We consider an example in the punctured plane, with $X$ given by the standard rotation generator $ X = y \pa_x - x \pa_y $ and (writing for ease of notation $r^2 = x^2 + y^2$) with $\mu$ given by 
$$ \mu = (-y/r^2) \d x + (x/r^2) \d y \ . $$ 
This is singular for $r=0$, and now the domain on which $\mu$ is a proper form is $B_0 = \R^2 \backslash \{ O \}$, which has a nontrivial cohomology.

Writing $\theta = \arctan(y/x)$, this corresponds to $$ \ga \ =\ \exp\[\arctan(y/x)\] \, = \, \exp[\theta] \ , $$ and then to the vector field $\XT = [\exp(\theta)] \pa_\theta$. Note that here $\ga$ is well defined only locally, as it corresponds to a multivalued function. One can check that $$ \zeta_1 \ := \ e^\theta u_\theta \quad {\rm and} \quad \zeta_2 \  := \ e^{2 \theta} (u_{\theta \theta} + u_\theta ) $$ are invariant under the $\mu$-prolongation $Y$ of the vector field $X$. In the $x,y$ coordinates (but retaining the notation $\theta := \arctan (y/x)$ for ease of writing) these read 
$$ \begin{array}{l}
\zeta_1 \ = \ (\exp \theta)\ (x u_y - y u_x) \ , \\
\zeta_2 \ = \ (\exp(2 \theta)) \ \( y^2 u_{xx} + x^2 u_{yy} - 2 x y u_{xy} - x u_x - yu_y + x u_y -y u_x \) \ . \end{array} $$
Let $\zeta_3$ be any smooth nontrivial function $\zeta_3 = \zeta_3 (r, u_r , u_{rr})$. Then any PDE of the form $$ \De \ :=\ F ( \zeta_1 , \zeta_2 , \zeta_3 ) \ = \ 0 $$ is a second-order equation invariant under the $\mu$-prolongation  $Y$. Its symmetry reduction gives an ODE of the form $$\^\De \ = \ \^F(r,w_r,w_{rr}) \ := \ F (0,0,\zeta_3) \, = \, 0  $$ for the function $u=w(r)$.

\subsubsection*{Example 3}

A variant of the previous example is obtained considering an equation of the form $\Delta := F( \zeta_1 , \zeta_2 , \zeta_3) = 0$ where, using again polar coordinates $r,\theta$ for notational convenience, 
$$ \zeta_1 := r u_r \ , \  \zeta_2 := r u_r \log r - 2 i u_\theta \ , \ 
\zeta_3 := r^2 u_{rr} \ . $$
which contains also imaginary terms. This admits the scaling vector field $X = x \pa_x + y \pa_y = r \pa_r$ as a $\mu$-symmetry, with
$$ \mu \ = \ {i\over 2} \( (-y/r^2) \d x \, + \, (x/r^2) \d y \) \ . $$ 
As in Example 2, we have a singularity in the origin, but here the
$\mu$-symmetry corresponds to a standard symmetry $\~X$ via a {\it
double-valued} function $\gamma$; we have indeed $\~X = \exp (i \theta / 2) X$.

\subsubsection*{Example 4}

Consider the  KdV equation for $u=u(x,t)$ 
$$ \De \, := \, u_t + u_{xxx} + u \, u_x \ = \ 0 \ , $$ 
and the vector field 
$$ X \ = \ {1\over t} \, {\pa\over{\pa x}} \, + \, 
{1\over x} \, {\pa \over{\pa t}} \ . $$ We consider $\mu = (1/x) \d x + (1/t) \d t$; this corresponds to $\ga  = x t$.

Applying the $\mu$-prolongation $Y$ of $X$, we get 
$$ \begin{array}{rl}
\De^{(1)}  := & Y(\De)|_{\De=0} \, = \, (-2/ \ga) \, u_{xxx} \ , \\
\De^{(2)}  := & Y(\De^{(1)})|_{\De^{(1)}=0} \, = \, \rho \
u_{xxx}|_{\De^{(1)}=0}\, =\, 0 \ , \end{array} $$ 
where $\rho$ is a function of $x$ and $t$. Therefore, $X$ is a partial $\mu$-symmetry of order $\ell=2$.

The set ${\cal S}$ of solutions of the system $\De=\De^{(1)}=0$ is given by $u=(x+c_1)/(t+c_2)$, where $c_1$ and $c_2$ are arbitrary constants; this includes the  solution $u=x/t$ which is invariant under the $\mu$-prolongation of $X$.

As stated by remark 12, the set of solutions ${\cal S}$ turns out to be a globally invariant symmetric set of solutions under $\XT=\ga \cdot X$, which is indeed a standard partial symmetry for the KdV equation \cite{CGpar}.

Finally, if one looks for the reduction of the KdV equation under the above partial $\mu$-symmetry $X$, it is easy to see 
that the invariance condition $Q=0$ gives the invariant variable $z=x/t$, and substituting $u=w(z)$ into the equation one obtains 
$$ w_{zzz} \ + t^2 (  \ w \, w_z\ -\ z\, w_z\ ) \ = \ 0 $$  
which, as expected, is not a ``pure'' ODE for the unknown $w(z)$, but has the form (3.5) (the role of the variable $s$ is played here by $t$). Solutions invariant under the $\mu$-prolongation of $X$ are provided by the system $ w_{zzz} = w_z (w-z) = 0$, and from this one obviously recovers the solution $w=z=x/t$ found above.

\subsubsection*{Example 5.}

As another example, consider the Boussinesq equation 
$$\De\, :=\, u_{tt}+u_{xxxx}+u\, u_{xx}+u_x^2\, =\, 0 $$ 
which admits the vector field 
$$  X \, = \, {\pa \over \pa x} + {1 \over {t^2}} {\pa \over \pa t} - \( {2 x \over {t^2} } + {10 t \over 3} \) {\pa \over \pa u} $$ 
as a partial $\mu$-symmetry (of order $\ell=3$) with $\d \mu = (2/t) \d t$ (i.e. with $\ga=t^2$). One easily obtains indeed 
$$ \begin{array}{rl} 
\De^{(1)}:= & Y(\De) \ = \ (1/\ga) \, ( -10  t - 3 u_x - 2 t u_{xt} - {5\over 3} t^3 u_{x  x} - x u_{x  x})  \\ 
\De^{(2)}:= & Y(\De^{(1)})|_{\De^{(1)}=0} \ = \ \rho \ (  2 + u_{x t} + t^2 u_{x x}) \end{array} $$ where $\rho=\rho(t)$. 
The most general solution of the equation 
$\De^{(2)} = 0$ is 
$$ u\, = \, F(t) + G \( x - {t^3\over 3} \) - 2tx \ , $$ 
where $F,G$ are arbitrary functions of the indicated arguments. 
Substituting this in the Boussinesq equation gives, after some calculations, the general solutions 
$$ u(x,t)= -{t^4\over 3} - 2t x +2 c_1 t - {12\over (x - t^3/3-c_1)^2} \ {\rm and} \ u(x,t)=- {t^4\over 3} + c_2 t - 2t x + c_3 \eqno(6.1)$$ 
where $c_1,c_2,c_3$ are arbitrary constants.

Looking now for the solutions of the Boussinesq equation which are invariant under the $\mu$-prolongation of $X$, one obtains from the invariance equation $Q=0$, 
$$u(x,t) \, = \, w(z) -2zt -t^4 \ ; \ z := x- (t^3/ 3) \ , $$ 
and the Boussinesq equation becomes 
$$ (w_z^2 +w  w_{z z} + w_{zzzz})  - 2t(3  w_z  +  z w_{z z}) \ = \ 0 \ , $$ 
which is of the form (3.5), as expected. 
Solving the system of the two ODEs appearing in the parentheses in the above equation gives the same solutions obtained above (6.1) but with the restrictions $c_1=c_2=0$.

\subsubsection*{Example 6.}

In this and the following example we will consider two-dimensional systems of PDEs; in these we write $u^1=u(x,y)$ and $u^2=v(x,y)$.

It is not difficult to verify that the system 
$$ \cases{ 
u_y + x^2 v_x - y v_y \ = \ 0 \ , & \cr 
x u_x - y u_y + x^2 v_x^2 + y^2 \ = \ 0 \ . & \cr} \eqno(6.2) $$ 
or more in general, any system of equations of the form $\zeta_\a (y, v_y, x v_x, u_y + x^2 v_x, x u_x + x^2 y v_x) = 0$, admits the vector field $$X\, =\, x{\pa\over \pa x}$$ 
as a $\mu$-symmetry with $\mu = \Lambda_x \d x + \Lambda_y \d y$ and 
$$ \Lambda_x = \pmatrix{0 & y \cr 0 & 0 \cr} \ ; \ \Lambda_y = \pmatrix{0 & x \cr 0 & 0 \cr} \ . $$ 
In this example, the matrix $\ga$ is  given by 
$$ \ga \ = \ \pmatrix{1 & xy \cr 0 & 1 \cr}$$ 
and it is easy to verify that all systems $\zeta_\a=0$ are indeed symmetric under the standard symmetry in evolutionary form 
$$ \XT := \ \ga \cdot X \  =\ \( x u_x + x^2 y v_x \) {\pa \over {\pa u} } + x v_x {\pa \over \pa v} \ . $$ 
The reduction, imposing $Q^a=0$, i.e., $u_x=v_x=0$, gives $u=u(y),\, v=v(y)$, and the above system becomes a system of equations $\zeta_\a (y, v_y, 0, u_y, 0) = 0$ involving $u,v$ with their derivatives with respect to $y$ only.

In particular, the system (6.2) is reduced in this way to 
$$ \cases{ u_y - y v_y \ = \ 0 \ , & \cr  - y u_y  + y^2 \ =\ 0 \ , & \cr} $$ 
with solution $u=y^2/2+c_1$, $v=y+c_2$.

\subsubsection*{Example 7.}

Consider the system 
$$ \cases{ 
u_x \ = \-v_x \log (|v_x|) + v \ , & \cr 
v_x \ = \ 2 v_y - y^2 + u_y + (v_x - v_y)^2 \ , & \cr} $$ 
and the vector field 
$$ X \ = \ {\pa \over \pa x} + v {\pa \over \pa u} $$ 
with $\mu=\Lambda_x\d x+\Lambda_y\d y$ and 
$$ \Lambda_x = \pmatrix{1 & 0\cr 0 & 1 \cr} \ ; \ 
\Lambda_y = \pmatrix{0 & 0\cr 0 & 1 \cr} \ . $$ 
The matrix $\gamma$ is then 
$$ \gamma \ = \ \pmatrix{\exp x & 0\cr 0 & \exp(x+y) \cr} \ . $$ 
Direct computation shows that $X$ is a partial $\mu$-symmetry of this system: indeed, according to definition 5$'$, the first application of the $\mu$-prolongation does not give zero but produces the new system $$ 0 \, = \, 0 \ ; \ v_x \, = \, v_y $$ 
and one needs another application of the $\mu$-prolongation of $X$. Imposing $Q^a=0$ we get $u_x=v$, $v_x=0$; i.e. $v=v(y)$ $u=w(y)+ xv(y)$. Thus the reduction yields from the second equation of the system (the first one is satisfied) 
$$ w_y + x v_y + v^2_y + 2 v_y - y^2 \ = \ 0 \ , $$ 
which has the form (3.5), as expected in the case of partial $\mu$-symmetries; this forces $v_y=0$ and leads to the solution $u = y^3/3 + c x$, $v=c$, with $c$ an arbitrary constant.

\subsubsection*{Example 8.}

Let us consider the Euler equation 
$$ u_t \ + \ u \, u_x \ = \ 0 \ . $$ 
It is shown in \cite{GaMo} that this admits as $\mu$-symmetry, with 
$$ \mu \ := \ u \, \d x \ - \ (u^2 / 2 ) \, \d t \ = \ \a\ \d x + \b\ \d t \ , $$ 
the vector field 
$$ X \ = \ \exp[- (u^2/2) t ] \ \( [B(u) - A(u) t / u] \, 
{\pa \over \pa t} \ + \ A(u) \, {\pa \over \pa u} \) $$ 
where $A(u), \, B(u)$ are arbitrary functions. Note that for this $\mu$ the compatibility condition $D_t \a = D_x \b$ is satisfied only on the solution manifold $S_\De$.

The $\mu$-symmetry $X$ corresponds to a nonlocal ordinary symmetry $\XT$ of exponential type, 
$$ \XT \ := \  e^{\int (u \d x - (u^2 /2 ) \d t ) } \ X \ . $$

\subsubsection*{Example 9.}

Let us consider the Calogero-Degasperis-Ibragimov-Shabat equation 
$$ u_y \ = \   u_{xxx} \ + \ 3 \, u^2 \, u_{xx} \ + \ 9 \, u \, u_x^2 \ + \ 3 \, u^4 \, u_x ; $$ 
motivated by \cite{SeS} we will choose the horizontal form 
$$ \begin{array}{l} 
\mu \ = \ \a \, \d x \ + \ \b \, \d y \\ 
\a \ := \ u^2 \ \ ; \ \ \b \ := \ 2 u u_{xx} + 6 u^3 u_x + u^6 - u_x^2 \ . \end{array} $$ 
This does not satisfy (1.7) in the full jet space $J^{(3)} M$; on the other hand, (1.7) is satisfied on the solution manifold $S_\De$.
Indeed, on $S_\De$ we have formally $\mu = D P$ with $P = \int u^2 \d x $, as readily checked by explicit computation. Obviously $D_x P = \a$. As for $\b$, we have on $S_\De$ 
$$ D_y P \ = \ \int 2 u u_y \d x \ = \ \int \[ 2 u u_{xxx} + 6 u^3 u_{xx} + 18 u u_x^2 + 6 u^5 u_x \] \ \d x \ ; $$ 
integrating by parts this is just $\b$ given above.

The vector fields 
$$ X_1 := \pa_x + u_x \pa_u \ \ {\rm and} \ \ X_2 := \pa_y + u_y \pa_u $$ 
turn out to be $\mu$-symmetries for the CDIS equation with the $\mu$ given above\footnote{Actually, any $\mu$-symmetry (with the $\mu$ considered here) vector field $X = \xi (x,u^{(1)}) \pa_x + \eta (x,u^{(1)}) \pa_y + \phi (x,u^{(1)}) \pa_u$ for the CDIS equation is written as $X = h_1(x,y,u,u_x) X_1 + h_2(x,y,u,u_x) X_2$.}.

The nonlocal symmetries associated to these are $\XT_k  = \ga \cdot X_k$, and the function $\ga$ is (see section 4) $\ga = e^P = \exp [ \int u^2 \d x ]$. This is general a formal expression only, but a proper function when restricted to $S_\De$.

Note that both $X_1$ and $X_2$ have trivial characteristics; it is quite remarkable that nevertheless they originate a nontrivial nonlocal symmetry.

\subsubsection*{Example 10}

The Burgers equation $$u_{xx}-uu_x-u_t=0 $$ admits the vector field $$ X = (2 w_x + u \, w) {\partial\over{\partial u}} $$
where $w=w(x,t)$ is any solution of the heat equation $w_t=w_{xx}$, as a $\mu$-symmetry with
$$ \mu \ = \ {1\over 2}u \, \d x + \( {1\over 2} u_x - {1\over 4} u^2 \) \d t \ := \ \a \, \d x \ + \ \b \, \d t \ . $$
Clearly $D_t \a = D_x \b$ is satisfied only on the solution
manifold of the Burgers equation; therefore, this  $\mu$-symmetry is
equivalent to a nonlocal exponential symmetry. Actually, this nonlocal symmetry is also a potential symmetry \cite{BK}: this follows from the property of the Burgers equation of being a conservative divergence equation.

\vfill\eject

\end{document}